\RequirePackage[hyphens]{url}
\documentclass[titlecompat,tight,twocolumn,10pt,table]{article}
\usepackage{usenix}

\usepackage{xspace}
\usepackage{graphicx}
\usepackage{enumitem}
\usepackage{booktabs}
\usepackage{multirow}
\usepackage[binary-units=true]{siunitx}
\usepackage{soul}
\usepackage{amsfonts,amsthm,amsmath}
\usepackage{subcaption}
\usepackage{latexsym}
\usepackage{wasysym}
\usepackage{tcolorbox}
\usepackage{tikz}

\usepackage{siunitx}
\usepackage{framed}
\usepackage{bm}
\usepackage[ruled,noend]{algorithm2e}
\usepackage{algpseudocode}
\usepackage{tabularx}
\usepackage{pifont}

\newcommand\paraphraseme[1]{\textcolor{black}{\textit{#1}}\xspace}
\newcommand\anon[1]{\textcolor{black}{\textit{#1}}\xspace}

\newcommand\id[1]{}

\renewcommand{\paragraph}[1]{\vspace{5pt}\noindent\textbf{#1.}\xspace}
\newcolumntype{R}{>{\raggedleft\arraybackslash}X}
\newcolumntype{L}{>{\raggedright\arraybackslash}X}
\newcolumntype{P}[1]{>{\centering\arraybackslash}p{#1}}

\newcommand{\eg}{e.g., }

\newcommand{\ie}{i.e., }
\newcommand{\etal}{et al.\@\xspace}

\newcommand{\vs}{victim-survivor\xspace}

\newcommand{\vses}{victim-survivors\xspace}
\newcommand{\Vses}{Victim-survivors\xspace}

\newcommand{\helpparagraph}[1]{\vspace{5pt}\noindent\textbf{#1?}\xspace}

\begin{document}

\title{Understanding Help-Seeking and Help-Giving\\on Social Media for Image-Based Sexual Abuse}

\author{
{\rm Miranda Wei}\\
University of Washington, Google
\and
{\rm Sunny Consolvo}\\
Google
\and
{\rm Patrick Gage Kelley}\\
Google
\and
{\rm Tadayoshi Kohno}\\
University of Washington
\and
{\rm Tara Matthews}\\
Google
\and
{\rm Sarah Meiklejohn}\\
Google
\and
{\rm Franziska Roesner}\\
University of Washington
\and
{\rm Renee Shelby}\\
Google
\and
{\rm Kurt Thomas}\\
Google
\and
{\rm Rebecca Umbach}\\
Google
} 

\maketitle

\begin{abstract}
Image-based sexual abuse (IBSA), like other forms of technology-facilitated abuse, is a growing threat to people's digital safety. Attacks include unwanted solicitations for sexually explicit images, extorting people under threat of leaking their images, or purposefully leaking images to enact revenge or exert control. In this paper, we explore how people seek and receive help for IBSA on social media. Specifically, we identify over 100,000 Reddit posts that engage relationship and advice communities for help related to IBSA. We draw on a stratified sample of 261 posts to qualitatively examine how various types of IBSA unfold, including the mapping of gender, relationship dynamics, and technology involvement to different types of IBSA. We also explore the support needs of \vses experiencing IBSA and how communities help \vses navigate their abuse through technical, emotional, and relationship advice. Finally, we highlight sociotechnical gaps in connecting \vses with important care, regardless of whom they turn to for help.
\end{abstract}
\vspace{10pt}

\noindent\fbox{%
    \parbox{0.98\columnwidth}{%
        \textbf{Warning:} This paper includes descriptions and quotes about image-based sexual abuse. Such material is disturbing.
    }%
}
\section{Introduction}

People increasingly share sexually explicit images with consent in intimate relationships~\cite{sexting-rates}, as cultural norms change and image sharing capabilities increase.
However, this trend has coincided with a rise in \emph{image-based sexual abuse} (IBSA): a continuum of harassment and scams involving the receipt, generation, and distribution of sexually explicit images~\cite{mcglynn2017image, mcglynn2017beyond}. Examples include unwanted solicitations for sexually explicit images~\cite{ringrose2021understanding, sparks2023image}, sextortion~\cite{cross2023pay, o2022cyber}, and nonconsensually sharing sexually explicit images~\cite{mcglynn2017beyond}. In terms of scale, one in ten women under the age of 30 in the US has been threatened with or experienced the nonconsensual sharing of their nude images~\cite{lenhart2016nonconsensual}, and roughly one in twenty adult men in the US has experienced sextortion~\cite{eaton2023relationship}.

As with other forms of technology-facilitated abuse---including intimate partner abuse~\cite{tseng2020toolsIPV}, stalkerware~\cite{freed2018stalker}, and interpersonal surveillance~\cite{wei2022tiktok}---the support needs of people experiencing IBSA (\vses) are complex. Perpetrators can be intimate partners, peers, or strangers.  Even against perpetrators with basic technical capabilities, preventing the distribution of sexually explicit images can be daunting. Support available today includes image fingerprint databases used by platforms to take down imagery~\cite{ncmec-takedown, stop-ncii} and institutional guides on how to respond to IBSA~\cite{ncmec-sextortion}. However, \vses may be unaware of these resources or find them ineffective, leading them to seek alternative support.

In this paper, we explore how adults seek and receive help for IBSA on Reddit, a popular social media platform for threaded dialogue. Given the scarcity of \vses who turn to law enforcement or platforms for help~\cite{campbell2022social, ruvalcaba2020nonconsensual}, social media provides an important avenue for disclosure and support. Expanding on knowledge from prior work studying help-seeking on Reddit for sexual abuse (e.g., rape) \cite{andalibi_2016, moors2013dance, oneill2018today}, we focus on help-seeking on Reddit across the continuum of \textit{image-based sexual abuse}.
Specifically we investigate:

\begin{enumerate}[label={\bfseries RQ\arabic*:},leftmargin=1cm, topsep=0pt,itemsep=0ex,partopsep=1ex,parsep=1ex]
    \item \textbf{IBSA types}. What types of IBSA do people seek help for on Reddit? 
    What gender and relationship dynamics between perpetrators and \vses do they disclose? How might differences between IBSA types influence the development of supportive solutions?
    
    \item \textbf{Help-seeking}. What help do they seek? 
    How do their needs vary across IBSA types? What actions do they share that they have already taken?

    \item \textbf{Help-giving}. What help do they receive? 
    How supportive is it?
    What gaps does this help fill compared to other interventions? What gaps remain?
\end{enumerate}

To answer these questions, we used a mixed methods approach to sift through 5.7 million English-language Reddit posts shared on relationship and advice subcommunities over the last 3 years.
Leveraging a novel large language model (LLM) data processing pipeline and extensive manual review, we identified more than 100,000 queries for help related to IBSA---roughly 2\% of posts on the subcommunities. 
This method allowed us to analyze a much larger sample than previous qualitative Reddit work.
We drew on a stratified sample of these posts to qualitatively explore the continuum of IBSA including financial and nonfinancial sextortion, nonconsensual \textit{synthetic} explicit imagery, pressurized sexting, cyberflashing, nonconsensual explicit imagery, and recorded sexual assault. 

We found that although IBSA covered a wide range of contexts, perpetrators, and harms, \vses nevertheless shared similar needs: to be heard and supported through life-changing experiences of abuse.
Across types of IBSA, \vses sought information about their options (technical, legal, or otherwise), advice for coping with distressing emotions, and suggestions for managing relationships. Timely support was crucial: in half of the cases we analyzed, \vses were seeking immediate help for active IBSA.
Though the help provided by the Reddit community was sometimes oversimplified and made limited use of institutional support resources, it was also largely empathetic and validated \vses' experiences, helping to partially address their needs.

Our work characterizes IBSA to chart additional directions to support \vses, in ways that complement the distinct avenues for help that exist today.
We reflect on the role of technology in facilitating IBSA and identify opportunities for technologists and platforms to help prevent or mitigate IBSA, while also discussing how our insights can inform advocates in providing support for \vses.

\section{Related Work}

\subsection{Image-based sexual abuse (IBSA)}
\label{rw:ibsa}
IBSA is an umbrella term referring to ``taking, distributing, and/or making threats to distribute, nude or sexual image[ry]\footnote{As with prior work~\cite{powell2019image}, we use  ``imagery'' to include photos and video.} without a person’s consent''~\cite[p. 1]{powell2019image} (see also:~\cite{dekeseredy2016thinking, mcglynn2017image, mcglynn2017beyond, powell2017sexual, powell2018image}). For the purposes of our study, we expand this definition to include the unwanted receipt of and synthetic generation of sexually explicit\footnote{Sexually explicit imagery can be distinguished from intimate imagery, which also includes images of people in private or sensitive contexts (e.g., sleeping, in states of intoxication, or without religious coverings).} 
images, that is, images portraying nudity or sexual conduct.
IBSA is a type of technology-facilitated abuse in which a person leverages technology to exert control over another~\cite{bluett2013role, dekeseredy2016thinking, henry2020technology}. We refer to the person enacting IBSA as the \emph{perpetrator} and the person experiencing IBSA as the \emph{\vs}.
 
\paragraph{Continuum of IBSA}
IBSA exists on a continuum, occurring in diverse relational contexts, with distinct motivations, and rapidly evolving threats~\cite{mcglynn2017beyond}. The types of IBSA that inform our investigation include:

\begin{enumerate}[topsep=0pt,itemsep=0ex,partopsep=1ex,parsep=1ex]
\item \textit{Sextortion}: a perpetrator makes threats to distribute sexually explicit images of a person unless they comply with the perpetrator's demands~\cite{cross2023pay, canadian2022sextortion, henry2015beyond, o2022cyber}.

\item \textit{Nonconsensual synthetic explicit imagery (NCSEI)}:
a perpetrator uses software (e.g., photo editing or generative AI tools) to create sexualized depictions of a person (\eg ``deepfakes,'' ``cheapfakes'')~\cite{flynn2022deepfakes, henry2018ai, mccosker2022making, okolie2023artificial, winter2020deepfakes}.

\item \textit{Pressurized sexting (PS)}: a person experiences unwanted solicitation for explicit images (\eg ``coerced sexting'')~\cite{ringrose2021understanding, sparks2023image}. 

\item \textit{Cyberflashing (CF)}: a person receives an unwanted explicit image~\cite{mcglynn2021cyberflashing}. 

\item \textit{Nonconsensual explicit imagery (NCEI)}: a perpetrator uses explicit imagery for revenge or to enact control~\cite{mcglynn2017beyond} (\eg ``revenge porn''); or otherwise nonconsensually creates, retains, or distributes explicit imagery (\eg ``upskirting,'' ``downblousing'')~\cite{bell2006up, lewis2023upskirting}.

\item \textit{Recorded sexual assault (RSA)}: a perpetrator records or distributes imagery from a sexual assault~\cite{henry2018policing}.
\end{enumerate}

\paragraph{Role of technology}
Technology often plays a role in both the generation and distribution of IBSA. Outside of consensual sharing or nonconsensual recording, methods of obtainment include hacking~\cite{franklin2014justice}, photoshopping~\cite{mcglynn2019shattering}, and using generative AI tools to create synthetic imagery~\cite{mccosker2022making, umbach2024attitudes, winter2020deepfakes}. Distribution channels used by perpetrators include---but are not limited to--- websites~\cite{clancy2021just, henry2019image, maas2021slutpage, uhl2018examination}, social media platforms~\cite{franklin2014justice, mccosker2022making, semenzin2020use}, and mobile apps~\cite{huber2023image, mcglynn2021s}.

\paragraph{Harms}
Harms from IBSA---as with other forms of sexualized violence---are not uniformly experienced, but are nonetheless serious and consequential~\cite{bates2017revenge, bloom2014no, kamal2016revenge, ruvalcaba2020nonconsensual}, carrying emotional, physical, financial, and social impacts~\cite{huber2023shadow}. When images are distributed, victim-survivors become visually recognizable to friends, family, co-workers, or others~\cite{huber2023shadow}, and imagery is often shared with other personally identifiable information (e.g., names, social media handles, phone numbers, and/or addresses~\cite{citron2014criminalizing, franklin2014justice, stroud2014dark}). The often permanent nature of online material intensifies IBSA harms, as imagery can be repeatedly downloaded, saved, and shared, making ``complete" removal challenging, or even impossible~\cite{bartow2012copyright}. As IBSA often co-occurs with other forms of gender-based violence (e.g., intimate partner abuse, stalking, and sexual harassment)~\cite{mcglynn2021s}, harms may compound with other polyvictimizations.

\subsection{IBSA help-seeking and help-giving}
Overall, rates of IBSA help-seeking through peers and family, institutional support, and platforms are low~\cite{campbell2022social, ruvalcaba2020nonconsensual}. One potential reason is that in contrast to other forms of sexualized violence, targets of IBSA are inherently not granted anonymity, as they are facially (or otherwise) identifiable~\cite{huber2023shadow}. Another challenge is that help-givers may hold victim-blaming attitudes, which have been found among the general public~\cite{call2021perceptions, flynn2023victim, zvi2021perceptions}, law enforcement~\cite{bond2021understanding, zvi2020police}, and victim-survivors' friends and family~\cite{sparks2022snapshot}. 

Online help-seekers for topics such as health often look for specific advice or information, acknowledgement, or sympathy~\cite{andalibi_2017, greene2011online, newman_2011}. 
Prior work also studies online help-seeking for sexual abuse (e.g., rape); by contrast, our work considers online help-seeking for \textit{image-based} sexual abuse.
When disclosing online about sexual abuse (that was not image-based), \vses perceived online fora as safer spaces to disclose stigmatized experiences and connect with others, with reduced interpersonal risk~\cite{andalibi_2018, andalibi_2016, kamarudin2018study, moors2013dance, oneill2018today}.
As we show, social media---specifically Reddit---has at least hundreds of thousands of posts related to IBSA help-seeking. 
Thus it is crucial to understand this support system and how best to enmesh it with other social, institutional, and platform resources.

\begin{figure*}[t]
    \centering
    \includegraphics{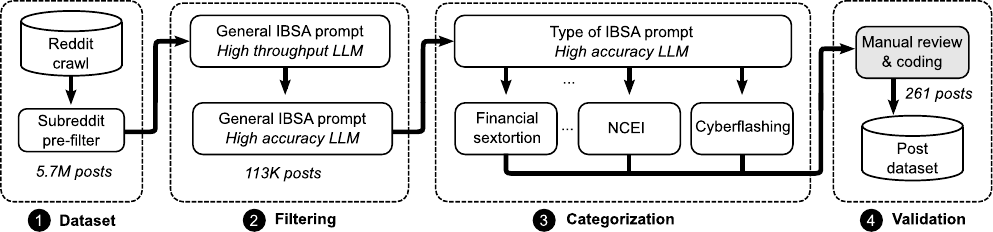}
    \caption{Data processing to identify help-seeking posts on Reddit. We relied on recent advances in LLMs to automatically sift through 5.7M Reddit posts to identify 113K posts likely related to help-seeking for IBSA. We automatically labelled each post with one of seven types of IBSA. Coders randomly pulled a sample of these posts to manually validate the LLM-generated metadata and reach a minimum threshold of 40 posts for each of the seven IBSA types. As a post could mention multiple types of IBSA, our \textit{Post dataset} includes a total of 261 posts.}
    \label{fig:pipeline}
\end{figure*}
\section{Methodology}
We conducted a mixed-methods study of IBSA help-seeking and help-giving experiences on Reddit. We discuss the process we followed to identify IBSA posts, our qualitative coding practices, and the ethics and limitations of our approach.

\subsection{Data processing}\label{sec:data}
We built a data processing pipeline that automatically identified help-seeking for several types of IBSA discussed on Reddit. We outline the full pipeline in Figure~\ref{fig:pipeline} and discuss each component below.

\paragraph{Dataset}
The first step in our data processing pipeline was to identify Reddit posts and subreddits relevant to our research (Figure~\ref{fig:pipeline}, \ding{182}). Our Reddit dataset originates from a continuous Internet-wide crawl of public URLs in a way that respects robots.txt and other rules for crawlers. In order to reduce the search space for IBSA posts, we focused on 43 popular subreddits related to scams, advice, relationships, dating, and sex---including \textsf{r/advice}, \textsf{r/askwomen}, \textsf{r/askmen}, and \textsf{r/sextortion}---that we identified from an initial manual exploration (see Appendix~\ref{sec:subreddits} for the full list). 
All of these subreddits had at least 1,000 members.
In total, this dataset includes 5.7M English-language posts from 2.9M unique users (\ie ``original posters'' on Reddit, posters in this work) published between April 1, 2020 and September 12, 2023.

\begin{figure}[t] 
\centering

\begin{minipage}{\columnwidth}
   
   \begin{tcolorbox}[colback=green!5!white,
                    colframe=green!20!white,
                    arc=4mm,
                    ]
     \footnotesize {\bf Title:} [TITLE]\\
{\bf Article:} [ARTICLE]\\

{\bf Question:} Does the title or article discuss any form of image-based sexual abuse, such as sending or receiving unwanted nude images, unwanted flashing, being coerced into sharing nude images, fear of having nsfw or intimate images leaked or hacked, sextortion, nonconsensual image sharing, or revenge porn? Provide both an answer and a summary explaining your reasoning.\\
{\bf Answer:}
  \end{tcolorbox}
\end{minipage}
\caption{Initial prompt designed to identify posts that contained any discussion of IBSA.}
\label{fig:ibsa_filter_prompt}
\end{figure}

\begin{table}[t]
\centering
\begin{tabularx}{\columnwidth}{Xcrr}
\toprule
\bf Type & \bf Key & \bf Posts & \bf Threads \\
\midrule
Financial sextortion & FS & 50 & 22 \\
Nonfinancial sextortion & NFS & 46 & 26 \\
Nonconsensual \textit{synthetic} explicit imagery & NCSEI & 51 & 32\\
Pressurized sexting & PS & 45 & 34 \\
Cyberflashing & CF & 41 & 33 \\
Nonconsensual explicit imagery & NCEI & 45 & 35\\
Recorded sexual assault & RSA & 40 & 20\\
\midrule
\textbf{Total} (Any type of IBSA) & -- & 261 & 160 \\
\bottomrule
\end{tabularx}
\caption{Our qualitative dataset consists of seven distinct types within the continuum of IBSA, as defined in Section~\ref{rw:ibsa}. Prior work largely collapses FS and NFS into the single category \textit{sextortion}. For our purposes, the differences were meaningful, so we split them into two distinct categories. For each type, we include the total number of stratified posts and comment threads that we analyzed. A single post could describe experiences with more than one IBSA type.}
\footnotetext{Prior work largely collapses FS and NFS into the single category \textit{sextortion}. For our purposes, the differences were meaningful, so we split them into two distinct categories.}
\label{tab:ibsa-categories}
\end{table}

\paragraph{Filtering}
We analyzed each Reddit post using a suite of LLMs and prompts to identify those likely to be about help-seeking for IBSA (Figure~\ref{fig:pipeline}, \ding{183}). While we initially explored a keyword-based search, the terminology and context proved too nuanced to capture without a prohibitive number of false positives to manually review. Instead, we queried Google Cloud's lighter-weight Vertex AI \textsf{text-bison} LLM~\cite{vertex-ai} with a prompt to identify posts generally discussing IBSA (Figure~\ref{fig:ibsa_filter_prompt}), then refined the search by repeating the prompt with a more accurate (but expensive) \textsf{text-unicorn} LLM~\cite{vertex-ai}. 

We validated this approach using a manually curated test set of 80 posts that discussed IBSA (generated from our initial manual exploration and when exploring the feasibility of keyword matching), and 197 non-IBSA posts. Our chained prompt correctly identified 79 of the 80 IBSA posts (98.8\% recall) and 194 of the 197 non-IBSA posts (98.4\% specificity). While the final dataset of posts we use is manually validated to remove all errors, this high degree of recall and specificity gave us confidence that we did not incorrectly omit some IBSA concepts from our study, while at the same time reducing the toil of manual validation. When applied at scale, our filtering identified 113K Reddit posts likely discussing IBSA---2\% of posts on the subreddits we analyzed. 

\paragraph{Categorization} To ensure that our study captured distinct experiences in the continuum of IBSA, we queried \textsf{text-unicorn} to categorize each post by the type(s) of IBSA involved in order to support stratified sampling (Figure~\ref{fig:pipeline}, \ding{184}). 
We derived these types---listed in Table~\ref{tab:ibsa-categories}---based on our preliminary analysis of prior literature (Sections~\ref{rw:ibsa} \&~\ref{initialcodes}).
We found that the LLM struggled with some concepts more than others. For example, \textit{sextortion} was easy for the LLM to categorize, while \textit{recorded sexual assault} was more error prone (and also more rare). To counteract this and account for posts that discuss multiple types of IBSA, we also created (more computationally expensive) per-type prompts with multi-stage reasoning to distinguish overlapping concepts (\eg financial and nonfinancial sextortion). See Appendix~\ref{sec:extra-prompts} for the general categorization and specialized prompts. We rely on manual review to validate these labels, discussed next. Given the potential for error with automated categorization (absent manual review), we avoid providing a relative breakdown of the types of IBSA at-scale, and focus only on our stratified sample.

\paragraph{Validation} During validation, we selected from the outputs of each categorization prompt and manually vetted the labels (Figure~\ref{fig:pipeline}, \ding{185}). Here, we corrected for any missed or erroneous labels and discarded any posts that were not related to help-seeking for IBSA. We also removed posts where the \vs appeared to be under the age of 18.\footnote{Some posts may originate from minors, but we cannot confirm. The advice we analyzed appeared targeted towards adults. The nuances of help-seeking and giving are fundamentally different for child sexual abuse (\eg lack of agency, mandatory reporting) and thus beyond the scope of this work.} We repeated this until we achieved a minimum of 40 posts for each of the seven types of IBSA,\footnote{We sampled independently per IBSA type. Some posts discussed multiple IBSA types, which is why there are more than 40 posts for some types.} resulting in a final dataset of 261 posts. We sampled 40 posts per IBSA type to achieve meaning sufficiency~\cite{braun2022conceptual}, balanced with reasonable researcher effort. Some posts involved multiple types of IBSA, thus requiring we label fewer than 280 posts to meet our minimum sample size. Each post in our sample on average contained 285 words (min: 24, max: 2,363).

For each post, we re-crawled the live site for the current state of discussion (\ie all threads) on December 11, 2023. This yielded 2,159 threads consisting of 4,225 individual comments from 2,298 unique accounts. To allow for a robust qualitative analysis, we filtered this set down to the top three upvoted threads per post, and then randomly sampled 160 of these popular threads. We use \textit{posts} to understand the types of IBSA for which people were seeking help and what help they were seeking, and \textit{threads} to understand what help they received. We summarize our qualitative dataset in Table~\ref{tab:ibsa-categories}.

\subsection{Qualitative analysis}
\label{ss:coding}
We relied on a rigorous, qualitative analysis to describe and identify themes in our 261 posts and 160 threads.
We used codebook thematic analysis (TA) that combined inductive and deductive approaches, which aligned to our mixed-methods approach ~\cite{fereday2006demonstrating}. We employed a five-stage codebook TA \cite{roberts2019attempting}: (1) sourcing initial codes; (2) developing initial codes; (3) codebook design; (4) codebook application; and (5) interpretation. For this study, use of a codebook enabled a refined and focused analysis of qualitative data \cite{crabtree1999doing}.

\paragraph{Sourcing initial codes} 
\label{initialcodes}
As we developed our research questions, we conducted an initial analysis of IBSA literature and preliminary scan of  IBSA help-seeking posts on Reddit -- to summarize and identify potential codes \cite{crabtree1999doing} relevant to help-seeking needs, help-giving behaviors, and other characteristics of the data that applied to our research questions. Existing literature on IBSA types and risk factors informed our initial deductive codes; codes describing IBSA help-seeking needs and help-giving behaviors were developed inductively, as these are more emergent areas in the literature. 

\paragraph{Developing initial codes}
Next, one researcher familiarized themselves with the raw data, cumulatively reading thousands of Reddit posts about IBSA during the development of our data processing pipeline (Figure~\ref{fig:pipeline}). This researcher then applied the initial codes to a random set of 100 posts stratified across IBSA types and then generated new codes from the raw data to identify more useful attributes of the data. 

\paragraph{Codebook design} 
Next we developed the codebook, finalizing labels, definitions, and exclusions. All members of the research team reviewed the codebook; we took note of initial disagreements to iteratively update our codebook to account for these nuances.
We segmented our codebook into three sections. The first section focused on the \textit{nature of the IBSA}: the type of IBSA, the platforms involved, details about the perpetrator and \vs, the origin of the IBSA imagery, and method of distribution. The second section focused on \textit{help-seeking}: when help-seeking occurred, strategies already attempted, and what help was sought.
The third section focused on \textit{help-giving}: the type of support or advice offered and the interactivity of help-giving.
See Appendix~\ref{sec:codebook} for details on individual codes and definitions. 

\paragraph{Codebook application and reliability} Four coders applied the codebook to our posts dataset and three applied the codebook to our threads dataset, with two independently coding each post or thread. For reliability, we used \textit{consensus coding} \cite{cascio2019team} for consistent judgment \cite{boyatzis1998transforming}. All coders iteratively discussed disagreements via meetings or online chat. Between discussions, one coder reviewed remaining disagreements, resolved obvious or already-discussed issues, and noted where discussion was still needed.
We chose to ensure coding reliability through consensus coding and discussion because our codes were not mutually exclusive and the data were nuanced.

\paragraph{Interpretation} Lastly, we iteratively collated various codes, reviewed the data and memos, and discussed themes. Alongside our thematic analysis, we offer descriptive statistics. We report counts in the results with the notation \textit{X of Y}, where Y is the total number of posts, threads, or people about which that code is specified, that is, excluding unspecified. In sections where Y is constant, only X is reported for brevity.

\subsection{Ethics}
Similar to prior work~\cite{bellini2021narrative, tseng2020toolsIPV, wei2022tiktok}, we rely on data shared publicly by users on social media. We excluded all posts where it was clear that the poster/\vs was under the age of 18. Our work does not directly recruit participants, but our study plan was reviewed by experts at our institution in domains including ethics, human subjects research, policy, legal, security, privacy, and anti-abuse. While the institution of the authors who conducted the data analysis does not require IRB approval, we adhere to similarly strict standards.

To mitigate potential harms that may come from \vses' data being exposed to unexpected audiences, we rephrased all quotes to preserve meaning but obscure the original source. 
Strategies for disguising the source of content are increasingly common in research fields investigating social media data~\cite{dym2020fandom, proferes_2021}, as well as recommended by digital-safety researchers~\cite{bellini2024research}, particularly when obtaining informed consent is not feasible.
Scholars emphasize that disguise is an ethical practice for protecting participants' privacy~\cite{markham2012fabrication}; additionally, contacting the original posters of the content we studied could unnecessarily re-traumatize \vses.
After rephrasing quotes, we searched each rephrasing to ensure the original source was not identifiable in the returned results.
To balance protecting posters' privacy with data integrity, another researcher compared the rephrasing to the original quote to confirm the semantic meaning of the quote was not changed. This rephrasing was post-hoc and did not affect our analysis.

As this project involved sustained engagement with traumatic content, researchers involved in analysis took different measures to support their well-being, including: weekly individual and group check-ins, reading about secondary trauma~\cite{lipsky2009Trauma}, meeting with trauma-informed experts, having access to therapists, taking breaks (e.g., playing Tetris, which is being explored as a tool for reducing traumatic flashbacks~\cite{iyadurai2018Preventing, simons2021Tetris}), and restricting reading of traumatic posts to shared or designated workspaces. 
Our use of LLMs also reduced the amount of manual review required.

\subsection{Limitations}
As with all research, ours has limitations. Our crawl of Reddit may be incomplete, and our LLM-based search for IBSA help-seeking may miss some concepts. We attempted to minimize biases by validating our prompts on an independent sample of posts and by focusing on qualitative results rather than comprehensive quantitative findings. 
Additionally, our US-based research team---whose domain expertise includes computer security and privacy, human-computer interaction, criminology, gender-based violence, and social computing---apply our own interpretations to the stories shared on Reddit.

Our visibility into IBSA help-seeking and help-giving is also limited to what people mention when posting to Reddit in English.
Reddit users are predominantly in the US, but also in the UK, India, and Canada~\cite{reddit2024users}.
Additionally, comments studied in Section~\ref{sec:help-giving} do not include those removed by Reddit mods, whose invisible labor contributes valuable content moderation~\cite{li2022all}.
Percentages are reported in the results for reader ease but should not be interpreted as generalizable to all cases of IBSA, given the limitations of our data collection. 

\paragraph{How we report gender}
Because IBSA can be a form of gender-based violence, we coded gendered terminology about the victim-survivor(s) and perpetrator(s) when specified.\footnote{Gender was unspecified for 26\% of perpetrators and 42\% of \vses.}
One challenge inherent in analyzing social media data is that posts do not reliably or consistently provide gender disclosures.
We inferred gender in five ways: 
Reddit conventions to self-identify (e.g., ``21F'' meaning a 21-year-old female), gendered nouns (e.g., woman, boyfriend, girl), gendered pronouns (e.g., she, he), body parts (e.g., dick pic, breasts), or other (e.g., posting to a gendered subreddit, asking for opinions of ``other women'').
We most often inferred \vs and perpetrator gender through gendered nouns (24\%; 60\%) or self-identifications (56\%; 23\%), but also inferred based on solely pronouns (8\%; 17\%) and body parts (26\%; 1\%).
To avoid piecemeal reporting, we collapsed sex and gender, such that ``man'' includes all masculine terms, including gendered nouns and pronouns, and ``woman'' likewise includes all feminine terms. These codes should be interpreted as researcher inferences based on gendered terms in Reddit's broadly cisheteronormative context, which may or may not align with the gender identities of the individuals involved.

\section{Characterizing IBSA Experiences}
\label{sec:ibsa-types}

We begin by characterizing the IBSA experiences about which posters sought help. 
While prior work has identified the types of IBSA covered here, we contribute a description of all seven types from the same dataset, highlighting common, co-occurring, and distinguishing patterns across types. This approach provides an expanded understanding of a range of IBSA experiences and sets the stage for how to broadly support help-seeking needs.

In most cases, the poster identified as the \vs (in 237 of 261 posts; 91\%); in others (30; 11\%), the poster was seeking help on behalf of the \vs, \eg a friend or intimate partner.\footnote{Counts do not sum to 261 because some posters were both a \vs and seeking help on behalf of other \vses of the same perpetrator(s).} We synthesize these experiences, examining how abuse unfolded, perpetrators' apparent motivations, and gendered patterns between perpetrators and \vses. 

\subsection{Financial sextortion (FS)}
\label{ss:types-fs}
Financial sextortion occurred when a perpetrator threatened to expose explicit images of a \vs unless the perpetrator was paid money.

\paragraph{Clear methods to obtain images}
To initiate contact for financial sextortion, a perpetrator typically connected with a \vs via a dating or social media app before moving to direct messaging or a communication app, engaging in a conversation with the \vs within minutes to days. The perpetrator then coercively obtained, made claims about, or created explicit images following one or a combination of the following methods: they sent (inauthentic) explicit images to encourage the \vs to reciprocate; they claimed to have an explicit image of the \vs (with or without evidence); and/or they created an explicit image (e.g., by attaching the \vs's face---such as taken from a profile picture---to someone else's body; such synthetic images are further discussed in Section~\ref{ss:types-mm}).

Once the perpetrator had established or asserted that they had an explicit image, they issued a threat: either the \vs had to pay or the explicit image would be distributed (e.g., \anon{``Send me money or I'll share your nudes''}). Payment demands varied, but were typically the local equivalent of \$100--1500 USD (specified in 28 of 50 posts; 56\%).

\begin{table}[t]
    \centering
    \small
\begin{tabularx}{\columnwidth}{X|rrrrrrr}
\toprule
    \textbf{Perpetrator}
    &\rotatebox{90}{\textbf{FS}} & \rotatebox{90}{\textbf{NFS}} & \rotatebox{90}{\textbf{NCSEI}} & \rotatebox{90}{\textbf{PS}} & \rotatebox{90}{\textbf{CF}} & \rotatebox{90}{\textbf{NCEI}} & \rotatebox{90}{\textbf{RSA}} \\ 
\midrule
Stranger        &  48 &   18 &  31 &  12 &  26 &     6 &    8 \\
Intimate partner &   0 &   10 &   8 &  21 &   3 &    15 &    4 \\
Friend          &   1 &    0 &   5 &   5 &   4 &     7 &    8 \\
Ex-intimate partner         &   0 &   10 &   1 &   2 &   1 &     8 &    8 \\
Colleague       &   0 &    0 &   1 &   2 &   4 &     1 &    4 \\
Family member   &   0 &    0 &   1 &   1 &   1 &     2 &    0 \\
\midrule
Other           &   1 &    0 &   0 &   0 &   2 &     0 &    0 \\
Unspecified     &   0 &    8 &   5 &   3 &   0 &     6 &   10 \\
\bottomrule
\end{tabularx}
    \caption{Relationship of perpetrator(s) to \vs, as described in posts.}
    \label{tab:relationship}
\end{table}

\begin{table}[t]
    \centering
    \small
\begin{tabularx}{\columnwidth}{X|rrrrrrr}
\toprule
    \textbf{Perpetrator}
    &\rotatebox{90}{\textbf{FS}} & \rotatebox{90}{\textbf{NFS}} & \rotatebox{90}{\textbf{NCSEI}} & \rotatebox{90}{\textbf{PS}} & \rotatebox{90}{\textbf{CF}} & \rotatebox{90}{\textbf{NCEI}} & \rotatebox{90}{\textbf{RSA}} \\ 
\midrule
Woman           &  21 &    8 &  15 &   2 &   6 &     1 &    3 \\
Man           &   3 &   20 &  16 &  39 &  27 &    31 &   31 \\
\midrule
Ambiguous         &   4 &    2 &   3 &   0 &   0 &     0 &    0 \\
Unspecified &  22 &   15 &  16 &   1 &   8 &    13 &    6 \\
\bottomrule
\end{tabularx}
    \caption{Gender of the perpetrator(s), as described in posts. \textit{Ambiguous} refers to posts where the poster initially described the perpetrator with feminine gendered terms but shifted to masculine gendered terms when a false identity was revealed.}
    \label{tab:gender-perp}
\end{table}

\paragraph{Perpetrators were usually strangers, often women; \vses men} Of the 50 posts, 48 identified the perpetrator as a stranger (Table~\ref{tab:relationship}). Of the 28 posts that specified the perpetrator's gender, 21 were believed\footnote{Some posters, in line with prior work~\cite{canadian2022sextortion}, acknowledged that perpetrators may be using this gender identity as a false persona.} to be women (Table~\ref{tab:gender-perp}). Of the 27 posts that specified the \vs's gender, 26 were men (Table~\ref{tab:gender-vs}).
  
\paragraph{Fear \& self-blame}
Perpetrators tried to exploit the \vs's fear of embarrassment. Posters often emphasized the threatened scale of distribution: \anon{``I'll send these masturbation videos to everyone you know, including your friends, colleagues, and relatives.''} \Vses often appeared to be in a state of panic when they posted, and they sometimes partially blamed themselves, claiming that they \anon{``messed up''} or \anon{``feel so stupid.''}

\begin{table}[t]
    \centering
    \small
\begin{tabularx}{\columnwidth}{X|rrrrrrr}
\toprule
    \textbf{Victim-survivor}
    &\rotatebox{90}{\textbf{FS}} & \rotatebox{90}{\textbf{NFS}} & \rotatebox{90}{\textbf{NCSEI}} & \rotatebox{90}{\textbf{PS}} & \rotatebox{90}{\textbf{CF}} & \rotatebox{90}{\textbf{NCEI}} & \rotatebox{90}{\textbf{RSA}} \\ 
\midrule
Woman           &   1 &   13 &  11 &  25 &  18 &    27 &   26 \\
Man          &  26 &    9 &  22 &   3 &   2 &     3 &    9 \\
\midrule
Unspecified &  23 &   24 &  19 &  18 &  25 &    18 &    8 \\
\bottomrule
\end{tabularx}
    \caption{Gender of the victim-survivor(s).}
    \label{tab:gender-vs}
\end{table}

\paragraph{What they've done; what else could they do} 
Posters sought help to determine if the threat was credible, prepare for what to expect next, and ask for strategies to cope with their fear. When seeking help, \vses often mentioned technical protections they had already taken, such as blocking the perpetrator, reporting the perpetrator to the platform, or bolstering their social media privacy and security settings. They then asked the community what else they could do.

\subsection{Nonfinancial sextortion (NFS)}
\label{ss:types-nfs}
At a high level, nonfinancial sextortion was similar to financial sextortion except the threats were not financially motivated.
 
\paragraph{Personalized demands for more images or control}
Nonfinancial sextortion tended to be customized to the particular \vs and their situation. Perpetrators---who were often intimate partners---used the threat of exposing an explicit image to demand new images and/or to exert emotional, physical, or sexual control over the \vs: \anon{``he was obviously getting mad and eventually exploded and told me he'd send out my nude pics if I didn't have sex with him.''} 

\paragraph{Perpetrators often were intimate partners, usually men; \vses women}
Nonfinancial sextortion was usually perpetrated by a current (10 of 38 specified; 26\%) or former intimate partner (10; 26\%) (Table~\ref{tab:relationship}).\footnote{We coded posts about sextortion that did not mention a demand for payment as NFS. Such posts where the perpetrator was a stranger resembled (and may have been) FS, so we omit them from discussion here.} \Vses often noted that the image the perpetrator threatened to expose was obtained with consent in a different context (8 of 43 specified), or nonconsensual secret recording (4). When specified, perpetrators tended to be men (20 of 30; 67\%), \vses women (13 of 22; 59\%)  (Tables~\ref{tab:gender-perp} \&~\ref{tab:gender-vs}).

\paragraph{Fear \& helplessness}
Like financial sextortion, perpetrators of nonfinancial sextortion attempted to control \vses with fear of embarrassment. However, when compared to financial sextortion, posts about nonfinancial sextortion tended to include a more pronounced tone of helplessness: \paraphraseme{``I feel suffocated with him and I don’t see any type of future with him, but he has the power to ruin my life.''}

\paragraph{Relationship \& emotional advice} \Vses expressed uncertainty about how to navigate their relationship with the perpetrator: \anon{``should I cave to his demands so he doesn't dump me?''}
They also asked for help coping with the emotional trauma associated with this type of IBSA. 
\subsection{Nonconsensual \textit{synthetic} explicit imagery (NCSEI)}
\label{ss:types-mm}
Nonconsensual synthetic explicit imagery involved digitally manipulated explicit images.
A majority of cases co-occurred with financial sextortion (27 of 51 posts, 53\%; discussed in Section~\ref{ss:types-fs}). 
Among other cases, a main theme appeared to be about the perpetrator exploring fantasies, especially when the \vs and perpetrator were socially connected. 

\paragraph{Simple manipulations}
Posters explained that basic editing techniques were used to generate explicit images, using terms like \textit{``photoshopped''} or \textit{``filters with my face.''} 
In cases of financial sextortion, perpetrators often sent \vses a ``collage'' of explicit images and other media---a list of the \vs's social media connections, screenshots of messages between the \vs and perpetrator, or calls-to-action from the perpetrator to incriminate the \vs (\eg calling them a ``pedophile'')---to scare the \vs into paying.

In other cases, posters found or learned about explicit images on the perpetrator's---usually the poster's intimate partner's---device, presumably for personal use such as the perpetrator's sexual gratification or exploration. In these images, \vses' or other relations' faces were superimposed into explicit images: \anon{``I recently found out [my intimate partner] had tons of pictures with my friends and family where their faces had been photoshopped onto sexually explicit photos to simulate porn scenes.''} 

\paragraph{The emergence of sophisticated manipulations}
The (suspected) use of generative AI to create nonconsensual synthetic explicit imagery was mentioned in 3 of 51 posts (6\%). In those cases, \vses found the imagery's photorealism terrifying: 
\anon{``Someone used a nudifying app to make naked photos of me from my social media pictures. I'm terrified.''} 

\paragraph{Nonconsent, fear, \& disgust} 
The quality of the synthetic imagery was not a focus of posters; instead, they discussed the harm that even simple manipulations could create. 
In financial sextortion cases, \vses shared that synthetic imagery might deceive family or peers: \anon{``The nudes are fake, but it's not like my parents can tell that.''}
In other cases, posters expressed disgust because the synthetic imagery was nonconsensually created by people the poster knew, and the images portrayed people in sexually explicit ways that the poster found inappropriate or offensive.
For example, one poster found synthetic images of a friends' family member, which left the poster
\paraphraseme{``in tears... I'm utterly disgusted and distraught by the situation.''}
Another poster discovered that their partner had been creating images of their friends and family for years, and despite knowing they were fake, viewed the creation of the images as a grave breach of trust:
\paraphraseme{``The files... were created a few years ago and I cannot believe how blind I was. I trusted him and now I am broken.''}

\paragraph{Navigating relationships \& feelings}
Posters sought help for how to deal with perpetrators (including asking for relationship advice) and to make sense of how they felt about the discovery of nonconsensual synthetic explicit imagery.

\subsection{Pressurized sexting (PS)}
\label{ss:types-ps}
Pressurized sexting occurred when a perpetrator coerced or demanded that a \vs send explicit imagery.
In a majority of these posts (29 of 45; 64\%), \vses claimed they had not sent the images; rather, they asked for help mitigating coercion.

\paragraph{Navigating relationship expectations}
Pressurized sexting occurred primarily in relationships, that is, by current intimate partners of the \vs (21 of 42 specified; 50\%), or between strangers who were prospective partners via online dating or social media apps (12; 29\%). 

\paragraph{New vs.\ established relationships}
When pressurized sexting occurred in the early stages of a (potential) relationship, posters expressed feelings of being manipulated or ``used for nudes'': \anon{``I can’t believe someone asks for nude pictures, and their only motivation in talking to me is trying to get nudes.''} 

When in established relationships, \vses often attempted to resolve conflicts between what they were comfortable with and their intimate partners' expectations. For example, some asked if pressurized sexting warranted ending a relationship: \anon{``My boyfriend asks for nudes and if I refuse he ignores me until I give in. Can someone tell me if we can fix this, or if I should just get therapy to help me break up with him?"} This was especially true if the pressurized sexting was perceived as the sole problem in a relationship. \anon{``This guy is great except he asks me to send nudes... I don’t want to break up but I hate when he does this. What should I do?''} 

\paragraph{Perpetrators were usually men, \vses women} 
Most perpetrators were presumed to be men (40 of 43 specified; 93\%); \vses tended to be women (25 of 27 specified; 93\%) (Tables~\ref{tab:gender-perp} \&~\ref{tab:gender-vs}).

\subsection{Cyberflashing (CF)} 
\label{ss:types-cf}
Cyberflashing occurred when a \vs received unwanted sexually explicit imagery.

\paragraph{Online dating} Cyberflashing commonly occurred in the process of seeking romantic relationships online. \Vses frequently discussed whether they were out-of-step with shifting norms around sending or receiving explicit images: 
\anon{``Can someone please explain why people send unsolicited nudes?''}
In some cases, cyberflashing occurred as part of pressurized sexting (Section~\ref{ss:types-ps}), where a perpetrator would demand reciprocity after sending unwanted explicit images: \anon{``I told him I didn't want any, but he sent dick pics anyways...he asked for a picture of my breasts but I said no cuz I never send naked pictures of myself.''}
The tone of these posts was often that of frustration, disgust, and disillusion. 

\paragraph{Platonic relationships}Less often, cyberflashing occurred in platonic relationships (e.g., between coworkers or friends). In those cases, the poster was typically concerned about the ongoing relationship and if/how they should address the abuse.
 
\paragraph{Perpetrators were usually men, \vses women} Most perpetrators were men (27 of 33 specified; 82\%), while \vses were women (18 of 20 specified; 90\%).

\subsection{Nonconsensual explicit imagery (NCEI)}
\label{ss:types-ncei}
Nonconsensual explicit imagery involved perpetrators producing or distributing explicit images without the \vs's consent. 
It often involved abuse by an intimate partner, ranging from a one-time incident to ongoing abuse.

\paragraph{Perpetrators were usually men; often partners, friends, or peers. \Vses were usually women} 
\Vses were usually women (27 of 30 specified; 90\%),
while perpetrators were often men (31 of 32 specified; 97\%). Most perpetrators were known to the \vs (33 of 39 specified; 85\%) rather than strangers (6; 15\%).

\paragraph{Images were often recorded without consent}
The trusted status many perpetrators had with \vses enabled secret recordings (20 of 38 specified; 53\%).
\Vses shared how perpetrators recorded them: \anon{``took nude photos of me while sleeping''} and \anon{``recorded me [naked] in my room.''}
Less frequently, images were recorded through coercion (6; 16\%) or shared with consent in a different context (6; 16\%). 

\paragraph{Navigating relationships; fear of possession \& distribution}
Nearly half of the posts focused on how to navigate the relationship (e.g., whether and how to confront or break up with the perpetrator).
\Vses were concerned about the perpetrator's possession of the explicit imagery (18 of 41 specified; 44\%) or its distribution, \eg via messaging (14; 34\%), social media (9; 22\%), or public websites (3; 7\%). 
 
\Vses concerned mainly with \textit{possession} primarily sought help on how to negotiate with the perpetrator, delete any explicit imagery, know if the imagery had been shared, or know how many images existed: \anon{``How can I find and delete as many as possible of the naked photos he has of me?''} Conversely, \vses contending with \textit{distribution} focused their help-seeking on what recourse, if any, existed: \anon{I dunno what to do. He sent my mom my nudes because I was breaking up with him... Any advice would be appreciated.}

\subsection{Recorded sexual assault (RSA)}
\label{types-rsa}
Recorded sexual assault involved the creation or distribution of images of sexual assault. 
It frequently resulted in \vses experiencing ongoing sexual trauma in addition to trauma from images being created or distributed.

\paragraph{Perpetrators were often known men; \vses tended to be women}
Perpetrators were often men (31 of 34 specified; 91\%) that \vses identified as former intimate partners (8 of 32 specified; 25\%), friends (8; 25\%), colleagues (4; 13\%), or current intimate partners (4; 13\%), but also strangers (8; 25\%).
\Vses were often women (26 of 35 specified; 74\%), but sometimes men (9; 26\%).

\paragraph{Co-occurred with other abuse}
\Vses shared that their assault was often secretly recorded (23 of 40; 58\%), such as when they were unconscious, sleeping, drugged, or under the influence of a substance. Recorded sexual assault often co-occurred with other abuse (13 of 39; 33\%)---sometimes prolonged intimate partner abuse---leaving \vses in a state of severe trauma.
Most \vses shared that perpetrators retained recordings of the assault (26 of 40; 65\%); some perpetrators distributed the recordings via messaging (7; 18\%), social media (4; 10\%), or websites (4; 10\%). 

\paragraph{Trauma, processing, \& coping}
All of these posts were deeply troubling and resulted in trauma to the \vs and typically others who were exposed to the recording. Most \vses asked how to cope with or understand what they had experienced, as a precursor to processing their emotions or taking a step towards remediation: \anon{``Was this sexual assault?,''} \anon{``Did I just make poor decisions or was this wrong?,''} \anon{``I don’t know how to feel,''} and \anon{``I’m wondering if any part of this sexual assault was normal.''}

\subsection{Co-occurring IBSA types}
\label{ss:co-occurring}
54 posts described more than one IBSA type. Most prominently, of 51 posts on nonconsensual synthetic explicit imagery, 27 co-occurred with financial sextortion and 7 with nonfinancial sextortion, indicating that perpetrators commonly leveraged synthetic images to sextort. Cyberflashing and pressurized sexting co-occurred in 9 posts, indicating that perpetrators sometimes both sent and pressured the \vs for explicit images. Nonconsensual explicit imagery co-occurred with multiple other IBSA types in 9 posts (out of 45 total); these posts mostly described multiple abusive events.
One post described three types of IBSA (NCEI, cyberflashing, pressurized sexting), and one described four (NCEI, recorded sexual assault, pressurized sexting, nonfinancial sextortion).
\section{Help-Seeking for IBSA} 
\label{sec:help-seeking}
Across IBSA types, posts mentioned prior remediation efforts and included some common patterns of help-seeking questions, such as informational questions (Section~\ref{ss:informational}), as well as how to cope with emotions (Section~\ref{ss:coping-emotions}) or navigate relationships (Section~\ref{ss:relational})---these and other less common help-seeking types are summarized in Table~\ref{tab:help-sought}. 

\begin{table*}[t]
\small
\begin{tabular}{lp{8cm}rrrrrrr}
\toprule
    \textbf{Help Sought}
    & \textbf{Description}
    & \textbf{FS} & \textbf{NFS} & \textbf{NCSEI} & \textbf{PS} & \textbf{CF} & \textbf{NCEI} & \textbf{RSA} \\ 
\midrule
Informational & Seeking general advice about options or potential actions.
    & 37 & 25 & 29 & 9 & 22 & 20 & 16 \\
Therapeutic & Seeking advice about emotions or other distressing elements.
    & 17 & 11 & 16 & 19 & 10 & 7 & 25 \\
Relational & Seeking advice about interpersonal dynamics or managing relationships with others.
    & 3 & 17 & 13 & 25 & 14 & 18 & 5 \\
Legal  & Seeking legal avenues of recourse or asking legal questions.
    & 5 & 4 & 8 & 3 & 7 & 9 & 7 \\
Technical & Seeking advice about mitigating the abuse through technical means, potentially to prevent primary or secondary sharing of the image, or tracking the past sharing of the image.
    & 3 & 1 & 2 & 1 & 1 & 4 & 1 \\
    \midrule
Other & Seeking a type of help not listed above.
    & 4 & 3 & 2 & 1 & 0 & 0 & 1 \\
\bottomrule
\end{tabular}
\vspace{-0.5em}
\caption{Types of help sought across the seven types of IBSA. Some posts discussed multiple forms of IBSA and were counted in multiple cells per row; some posts also included multiple types of help-seeking.}
\vspace{-0.5em}
\label{tab:help-sought}
\end{table*}

\subsection{When posters sought help}
\label{sec:dfaru-states}
Using a theoretical framework~\cite{jigsawmedium}, we can describe patterns in \textit{when} people sought help for IBSA. This framework defines four states of users experiencing safety events: \emph{prevention, monitoring, crisis,} and \emph{recovery} (see details in Appendix~\ref{sec:codebook}). 
People may move through the states nonlinearly and experience multiple events at once. 
When coping with \emph{multiple safety events} (57 of 257 posts; 22\%)\footnote{Number of posts here is 257 because four did not specify the user state.}, posters' panic and trauma were compounded.

Help was most often sought while the poster was in \textit{crisis}, that is, when they were actively dealing with abuse (142; 55\%).
These posters expressed panic and needed immediate reassurance, as well as clear, simple guidance on the most critical steps to stop or limit further damage.
Many posters also sought help while \emph{recovering} from the abuse and feeling traumatized by it (112; 44\%).
Fewer posters described being in \emph{monitoring}, for example, watching for new abuse or for images to appear in new places (42; 16\%). 
Rarely, posters sought help \emph{preventing} victimization (5; 2\%).

\subsection{Prior attempts at remediation}

Posters mentioned prior attempts at remediation (in 176 of 261 posts; 67\%), which were only partially successful or failed, prompting them to seek help on Reddit  (Table~\ref{tab:strategies-attempted} in Appendix~\ref{sec:extra-results}).

\paragraph{(Dis)engaging perpetrators} Of posts that noted prior attempts at remediation, nearly half mentioned engaging with perpetrators through negotiation or mediation (76 of 176 posts; 43\%).
This strategy often occurred when IBSA was enacted by an intimate partner, including NCEI and pressurized sexting.
Another nearly half disclosed that the \vs disengaged from the perpetrator, cutting off communication (82; 47\%); some attempted engaging and then disengaging.
Disengaging was common when the perpetrator was a catfisher or scammer, including financial sextortion and nonconsensual synthetic explicit imagery.

\paragraph{Technical strategies} 
Technical strategies, which included securing accounts or devices, platform reports, and deleting content, were sometimes employed (49 of 176 posts; 28\%). 
They were infrequently reported across IBSA types, except for financial sextortion, for which \vses more consistently reported deleting accounts or reporting.
Often, sexually explicit images resided on a perpetrator's personal device or chat history, making it technically challenging for \vses to access and for platforms to respond.
As many help-seekers posed open-ended questions (e.g., \textit{``what should I do?''} as described in Section~\ref{ss:informational}), some may not have been aware of what technical strategies were possible.

\paragraph{Social \& institutional support}
\Vses rarely mentioned seeking or obtaining social support---a common practice used to cope with other forms of abuse~\cite{warford2021sok, sambasivan2019they}---or broader institutional support. The small number of posts that did, mentioned filing police reports (20 of 176; 11\%), reaching out to family or peers (19; 11\%), or reaching out to the \vses' workplace (6; 3\%).
A common reason for not seeking social or institutional support was embarrassment:  
\anon{``I only told a few friends about what happened, but not with any specifics because it’s so humiliating. I really have no one to discuss this with.''}

\subsection{Information: To understand \& stop abuse}
\label{ss:informational}

Nearly half of posters sought \emph{informational} help (in 127 of 261 posts; 49\%), with questions focused on making sense of the IBSA and determining how to make it stop.

\helpparagraph{What should I do}
Many posts included open-ended requests
(\anon{``someone help me, i have no idea what to do''}), such as from \vses who appeared panicked as they tried to stop or recover from IBSA. 

\helpparagraph{What's happening to me}
Another type of request---especially from those experiencing recorded sexual assault---asked whether their situation constituted assault, abuse, or coercion. These requests sometimes included misunderstandings, such as how intoxication affects consent: 
\anon{``Was I sexually assaulted? I was okay going to his place even though I was totally wasted. When I was there, he made me do things and I consent after he convinces me. Then he takes out his phone to take a video, which I never consented to. Did I make the wrong choice or was it just wrong?''}

\helpparagraph{What can I expect}
Posters also asked what to expect in the near and long term. For example, for financial sextortion, \vses asked if perpetrators would follow through on threats to distribute imagery, if they were ``safe'' after waiting a certain period of time, or how they would know if the abuse was ``over.''
For nonfinancial sextortion and NCEI especially, \vses asked about what would happen if the explicit imagery were shared.
They sought help anticipating and mitigating future harm.

\helpparagraph{Why did they do this}
Questions about perpetrator motivations were common (e.g., why they created synthetic explicit imagery, or pressured people into sexting).
Multiple cyberflashing \vses---especially women---asked why it was \textit{so} common in online dating: 
\anon{``Why is it that within a few days of talking, men always send pen*s pics?''}

\subsection{Therapeutic: Coping with emotions}
\label{ss:coping-emotions}
About a third of posters sought emotional or \emph{therapeutic} help (in 86 of 261 posts; 33\%).

\paragraph{Trauma} 
\Vses asked for help coping with intense feelings, particularly in cases of recorded sexual assault and NCEI.
They were overwhelmed by emotions like shame 
(\anon{``I’m melting from the shame and guilt''}) and disgust
(\anon{``I'm disgusted and feel so violated''}) upon discovering the details of the IBSA, such as who images had been shared with, or who possessed (and thus had control over) the images.

\paragraph{Empathy, self-doubt, \& isolation}
Posters sought empathy (\anon{``Can anyone provide reassurance or share their experience, anything helps''}) and help with processing their feelings (\anon{``I needed to vent to someone''}).
Some felt conflicted and experienced self-doubt in understanding their own experiences: 
\anon{``I don't know how to feel, I don't trust anything.''}
For many \vses, experiencing IBSA was an isolating experience. They described being ostracized from social connections:
\anon{``I only wrote this post because I wanted to talk to someone. I have no friends right now''} or not receiving support from institutions like their employer's human resources department, their school, or law enforcement.

\paragraph{Fear of confronting the perpetrator}
Posters recognized that confronting the perpetrator could result in even more harm and asked for help in managing fears while planning next steps:
\anon{``I'm scared he'll be mad and share my pics if I confront him.''}

\subsection{Relational: Navigating relationships}
\label{ss:relational}
About a third of posters sought \emph{relational} help (in 84 of 261 posts; 32\%) for navigating situations where the \vs was socially connected to the perpetrator. 

\paragraph{Romantic relationships}
Some \vses experiencing NCEI as part of romantic relationships struggled with whether or not they should stay with their perpetrator-partner.
In these posts, ``love'' was cited as a reason to stay, and \vses questioned their own behavior:
\anon{``We really love each other. However I also feel bad because he’s always spending time with me. Is he stressed out because we spend so much time together? Would it have been better if I broke up with him?''} Posters also cited the central role their intimate partner played in their lives (\anon{``he's my everything''} and \anon{``everything i've done was with his support''}).

\Vses experiencing pressurized sexting and cyberflashing---typically when the perpetrator was an intimate partner---also asked about staying in the relationship (e.g., was the perpetrator's behavior a \textit{``red flag''} or \textit{``deal breaker''}?). 
They also commonly asked for help with how to effectively confront the perpetrator, such as to ask them to stop, resist pressure, or negotiate for images to be deleted.

\paragraph{Nonromantic relationships} 
When the perpetrator was socially connected to the \vs, but not an intimate partner, such as a work colleague or friend, \vses asked how to manage the relationship moving forward: 
\anon{``My [22M] BFF [22F] flashed me and now won’t talk to me. What should I do?''}

\section{Help-Giving for IBSA}
\label{sec:help-giving}

We now turn to comments on posts about IBSA, that is help-giving. 
Help-giving appeared to be community-oriented, as 55 help-givers (out of 2,298 total) supported multiple posters in our dataset, and all but seven of these commented within the same subreddit. 
Help-givers and help-seekers generally did not overlap (or they used throwaway accounts); only one account in our dataset created a post \textit{and} commented on someone else's post.

\begin{table*}[t]
    \centering
    \small
\begin{tabularx}{\textwidth}{lXrrrrrrr}
\toprule
\bf Help Given & \bf Description & \bf FS & \bf  NFS & \bf  NCSEI & \bf  PS & \bf  CF & \bf  NCEI & \bf  RSA \\
\midrule
Informational     &  General details that explain a type of IBSA, a perpetrator's motives, or directing to a resource for more information. & 15 &   11 &  17 &  12 &  17 &    14 &    7 \\
Technical         & Advice for security and privacy, blocking, reporting, deletion, and whether to engage with a perpetrator's account.  & 9 &   11 &  12 &  12 &  11 &    11 &    6 \\
Relational        & Advice for navigating relationships, whether to inform family and friends, or whether to reason with perpetrators.  &  4 &    7 &  10 &  16 &  11 &    14 &    5 \\
Therapeutic        & Advice for taking time for yourself, not to be afraid, that it is not your fault, and that you are not alone.  & 6 &    8 &  11 &  14 &   9 &    13 &    9 \\
Institutional support  &  Advice around law enforcement, legal options, therapists, support centers, or human resources. & 2 &    2 &   4 &   3 &   9 &     9 &    7 \\
\midrule
Counterproductive &  Advice that cast blame on the \vs, or otherwise ignored the needs of a \vs. & 5 &    2 &   4 &   4 &   1 &     2 &    5 \\
Other             & Help-giving in other ways not described above.  & 0 &    3 &   0 &   1 &   0 &     4 &    2 \\
\bottomrule
\end{tabularx}
    \vspace{-0.5em}
    \caption{Types of help given across the seven types of IBSA. Some threads responded to posts discussing multiple forms of IBSA and were counted in multiple cells per row; some threads also included multiple types of help-giving.}
    \vspace{-0.5em}
    \label{tab:help-given}
\end{table*}

\subsection{Types of support and advice}
\label{sec:giving-types}
Across our 261 posts, help was almost always given: only 15 posts received no comments, and the median post received five comments authored by four distinct users. However, few help-givers asked for more details to inform or tailor advice (14 of 160 threads). We discuss the most salient forms of help-giving below (Table~\ref{tab:help-given} provides a summary; some posts and associated threads contained multiple IBSA types, so the row totals are greater than the denominators reported below). 

\paragraph{Informational} The primary form of help-giving was sharing information (72 of 160 threads; 45\%).
For financial sextortion, informational advice highlighted the scripted nature of the attack: \anon{``This scam is super common. You can’t do anything except block them (and stop sending d*ck pics to randos).''} For cyberflashing and pressurized sexting, threads explained the changing norms around sexting and the choices available to \vses (e.g., blocking, reporting). For NCEI and recorded sexual assault, threads instead focused on reinforcing that the experience was abuse: \anon{``Revenge porn is pretty much always considered illegal''} and \anon{``This is definitely sexual assault.''} Help-givers' tone when sharing information, however, was not always comforting:
\anon{``Even if you told me to imagine the most fucked up thing I could, I still wouldn't think of that''}.
Furthermore, help-givers rarely pointed the poster to other communities or resources (5 of 160 threads; 3\%).  We discuss these limitations more in Section~\ref{sec:help-giving-limits}.

\paragraph{Technical} Technical advice was common across most types of IBSA (52 of 160; 32\%).
Some threads focused on not engaging with the perpetrator (21) or blocking and reporting a perpetrator's account (26), e.g., in financial sextortion, pressurized sexting, or cyberflashing, where a perpetrator was a stranger or acquaintance.
A less prevalent alternative was telling the \vs to make their own account private to minimize contact (7). Such advice was mostly absent from IBSA involving an intimate partner, like nonfinancial sextortion, NCEI, and recorded sexual assault. Technical advice in these scenarios focused on how to delete photos if the \vs had access to a perpetrator's accounts or devices (5), how to check for backups (2), or how to find and record evidence of abuse (9): \anon{``Screenshot the convo in which your gf got the video before you delete it.''}
Advice to block or otherwise avoid perpetrators aligns with highly recommended advice by experts to stay safer online from harmful content~\cite{wei2023responsibility}.

\paragraph{Relational} Threads frequently discussed relationship advice (52 of 160; 32\%), but not always in well-reasoned ways.
Threads commonly directed \vses to end their relationship (29) across all types of IBSA other than financial sextortion, no matter the circumstances shared by the \vs, including their living or financial situation, their desire to make a relationship work (\anon{``I love him so much, and I know he loves me too''}), or whether or not leaving was a safe option. For example, one \vs clearly stated their desire to stay with their partner, who was not the perpetrator of the IBSA, but was still told to leave: \anon{``That's fucking me up to the point I would dump her.''} Furthermore, help-givers at times assumed worse behavior from the perpetrator than was specified in the post: \anon{``Any guy who pressures you for nudes will share them with his friends.''} Some threads were more nuanced, however, laying out how the \vs had agency in setting boundaries in cases of cyberflashing or pressurized sexting (20). Others advocated engaging with a perpetrator (10), particularly for NCEI and pressurized sexting, to negotiate deleting images or ceasing requests: \anon{``Send him a message asking him to delete what he filmed, that he didn't ask for permission and broke your trust.''} Finally, threads discussed creating a support network to prepare for the risk of leaked imagery or to intercede with the perpetrator (12).

\paragraph{Therapeutic} Therapeutic help-giving was also common (52 of 160; 32\%),
such as reassuring \vses that a situation was not their fault (15) and telling them to not be alarmed (14) or look after themselves (12).
Threads also reiterated that a perpetrator was harmful (8) or commiserated with similar experiences (8).
In the case of financial sextortion, threads focused on reassuring the \vs that not engaging was the right tactic and to not be afraid: \anon{``If you haven’t sent any money, they have no reason to go after you.''} Meanwhile, threads responding to NCEI, recorded sexual assault, and nonconsensual synthetic explicit imagery emphasized general support of the \vs as they navigated abuse: \anon{``I hope that you heal and find sweet and honest love again.''} 

\paragraph{Institutional support} While less frequent, help-giving also encouraged seeking institutional support (31 of 160; 19\%).
This advice focused on contacting law enforcement (15) or accessing legal advice (13), therapists (8), advocacy groups (3), human resources (3), or immediate medical support (1). Such advice was largely given for cases of NCEI (9), cyberflashing (9), or recorded sexual assault (7). 

\paragraph{Counterproductive} Not all help-giving was supportive of \vses (19 of 160; 11\%).
Some community members admonished the \vs for their behavior, asking what they expected by sharing explicit images with others: 
\anon{``Honestly if you’re sending dick pics to online randoms, does it matter that they’re getting leaked?''}

Others minimized or de-legitimized \vses, asking \textit{``What’s the worst that can happen?''} or claimed most recorded sexual assault videos found online are fake.  While the majority of upvoted threads were helpful (144 of 160; 90\%), these examples show how help seeking on social media can expose \vses to additional shame or risk.

\subsection{Victim-survivor reactions to help given}

In general, the advice given on Reddit seemed to resonate with the poster.  Of the 246 posts with at least one comment, at least 64 had one or more comments from the poster, that is, where the poster engaged in some way with help-givers.\footnote{This number is a lower bound: our quantitative analysis was based on our delayed crawl of the live site and thus could not identify engagement by a poster who had later deleted their account, which was a majority of posters.} The median engagement by posters was 50\% of threads.  

Posters expressed a range of reactions to help given, including providing or requesting more information (20 of 160; 13\%),
expressing appreciation for the help they received (16; 10\%),
and outlining a concrete plan of action based on the help they received (9; 6\%).
These reactions were spread roughly evenly across IBSA types. In many of these comments, the poster offered only a short thanks, even if there was no clear resolution. Some threads seemed to have helped them make sense of their experience: \anon{``Thanks for confirming what I thought.''} One back-and-forth sextortion thread helped the \vs better understand the nature of the scam and feel reassured as a result: \anon{``Thanks for the help and info, I learned something new.''} In very few cases (4; 3\%), 
posters expressed feeling hopeless even after receiving support: \anon{``It's just so rough that there's nothing we can do.''}

\section{Discussion}
We now reflect upon both the challenges in this space (Section~\ref{sec:help-giving-limits}) and possible roles of technology in solutions (Section~\ref{sec:discussion:techrole}). While our results provide insights into help-seeking and help-giving behaviors on Reddit, we argue that it is too early to speak definitively on the full set of challenges or solutions.
Thus, we encourage readers to view our discussion here as a results-informed exploration of the possible parameters of the challenge and solution spaces.

\subsection{Existing challenges}
\label{sec:help-giving-limits}

\paragraph{Limited nuance or formalized advice}
\Vses were offered emotional support in the form of concern and empathy (Section~\ref{sec:giving-types}), making it clear informal help-giving on Reddit provides a lifeline to those who might otherwise be unwilling or unable to disclose and seek help. This suggests findings from previous related work on online help-seeking for offline sexual abuse~\cite{andalibi_2018, andalibi_2016, gorissen, moore2001inferring, oneill2018today} are replicated for technology-facilitated sexual violence. Namely, the anonymity and visibility management often afforded by online spaces facilitates support seeking, both explicit and implicit. As in previous work~\cite{moors2013dance, webber2014sexual}, we also observed the constraints associated with online support, such as help-givers who defaulted to certainty around the situation or the ``right'' outcome without asking for more details or centering the expressed needs or desires of the victim-survivor, thus overlooking important nuances.
For example, the point at which a \vs leaves an abusive partner poses the greatest risk of harm to the \vs~\cite{tjaden2000extent}, so gender-based violence advocates typically scaffold the creation of a safety plan~\cite{campbell2009safety, davies2013domestic}.
However, help-givers commonly suggested leaving a relationship without mentioning safety plans or other risk mitigation strategies. Relative to other types of help given, referring to institutional support was the least common (Table~\ref{tab:help-given}). 
Formal organizations\footnote{Examples for tech-facilitated abuse more broadly include \url{https://www.ceta.tech.cornell.edu/} and \url{https://techclinic.cs.wisc.edu/}} may be better suited to providing long-term and holistic support, but it is clear that connecting \vses to support outside Reddit remains a challenge.  

On the other hand, our work contains hints of why \vses may have sought help on Reddit rather than via institutions.  Between the time at which our Reddit snapshot was collected and our crawl of the live site in December 2023, 99 posts and 146 poster accounts had been deleted. Further, 41 posts were from ``throwaway'' accounts, i.e., accounts not used again on Reddit, suggesting that posters valued Reddit's perceived anonymity or ephemerality.  We also observed in Section~\ref{sec:dfaru-states} that posters most often sought help when in active crisis, and may have viewed Reddit as a faster way to get help than finding and reaching out to an advocacy organization.  Of course, the fact that posters sought help on Reddit does not preclude them seeking help from other sources as well.

\paragraph{Underscoring challenges in technology-facilitated abuse}
Our analysis of IBSA help-seeking on Reddit exemplifies two key challenges from broader technology-facilitated abuse literature.
First, prior work emphasizes the weaponization of new technologies to expand the scope of potential targets~\cite{obada2022sok}.
For example, in most financial sextortion and cyberflashing posts in our dataset, the perpetrator was someone previously unknown to the \vs before the IBSA (Sections~\ref{ss:types-fs} and~\ref{ss:types-cf}); perpetrators leveraged new platforms and social discovery algorithms to find new \vses.
Another poster asked for preventative strategies against generative AI specifically, because they were concerned:
\anon{``with all the AI hype, many people around me are making nudes of others, especially of women.''} 
More generally, the imagery was synthetic in all NCSEI cases and \vses expressed distress over the ability of perpetrators to portray them in ways they had not consented to. 
Perpetrators will continue to leverage new technology, necessitating ongoing research to prevent its misuse. 
Second, our findings reiterate the occurrence of polyvictimization, as studied in technology-facilitated violence literature~\cite{henry2018review, marganski2022single, mitchell2018poly}.
Many \vses described experiencing other types of abuse that co-occurred with the IBSA.
Echoing research on intimate partner abuse~\cite{Freed-2019, tseng2020toolsIPV}, supporting \vses requires holistic and trauma-informed~\cite{chen2022trauma} approaches that do not regard technology as a panacea.

\paragraph{Aligning interventions with sites of harm}
While some technological interventions exist for different types of IBSA, our analysis raises challenges for their reach. For example, StopNCII~\cite{stop-ncii} helps \vses get images removed from social media platforms, but in only nine (of 45) NCEI cases did posters mention that images had been posted on social media. More often, posters were concerned about perpetrators nonconsensually retaining images (18 of 45) or distributing them via messaging apps (14 of 45) (Section~\ref{ss:types-ncei}). 
Generally, IBSA incidents in our dataset were most frequently carried out via dedicated communication apps or direct messages on social media (143 of 247, see Table~\ref{tab:platforms} in Appendix~\ref{sec:extra-results}).  
Fewer cases involved images being posted on one-to-many platforms: social media (72), porn websites (6), or unspecified websites (3). Nevertheless, reporting IBSA to platforms remains one of the few technical mechanisms to remove explicit images stored or shared on a platform, or to potentially take action against a perpetrator's account and prevent others from being targeted. 
Thus, while reporting can be an effective intervention in these cases, our findings suggest an additional need to create contextually specific technological interventions. 

\subsection{Role of technology in solutions}
\label{sec:discussion:techrole}

\paragraph{Expanding and integrating the support ecosystem}
Our analysis shows IBSA help-seeking on Reddit occurs in hundreds of thousands of posts, some with tens to hundreds of comment threads. This creates a burden for communities to triage and potentially leads to inconsistent advice depending on who responds in the moment. Recent advances in LLM agents could amplify help-giving: triaging incoming queries, requesting additional information as needed, providing guidance where best practices exist (\eg for financial sextortion), and connecting \vses with relevant advocacy organizations who provide hands-on support. Results from our study could guide the design of such an agent or other interventions, accounting for help-seekers' common emotional states, questions, and needs (Section~\ref{sec:help-seeking}). Such an agent may not replace the interpersonal therapeutic support that help-seekers received through other people on Reddit, but could be more consistently available while complementing other available help-giving pathways. Early research on a non-generative AI agent is underway,\footnote{See Umibot, \url{https://umi.rmit.edu.au}} but this remains a ripe area for exploration.

\paragraph{On-device detection and warnings}
Given the complexity of remediating IBSA \textit{after} an explicit image is shared, technology could play a \emph{preventative} role. Given the privacy sensitivities---particularly around messaging where much of IBSA occurs (Table~\ref{tab:platforms} in Appendix)---this technology would be best \emph{on-device} to minimize explicit images being sent to platforms. To this end, Apple recently announced opt-in, on-device detection of sensitive (\eg ``nude'') content~\cite{apple-nudity-detection}. While these technologies are in an early stage, they might take the form of nudges against sending explicit images (\eg pressurized sexting, financial sextortion); or blurring explicit images upon receipt (\eg cyberflashing). The effectiveness of these nudges---and expanding the detection capabilities to dangerous interactions (\eg early detection of sextortion patterns)---remain to be explored. Smartphones could also extend current alertness-detection methods (e.g., as used in face biometric systems to assess whether a person is awake) to prevent or make more challenging the taking of nude or explicit imagery of people who are asleep or otherwise unaware. 

\paragraph{Expanded controls around explicit content} Technology might also play a role in mediating how, for how long, and with whom explicit images are shared to reduce the risk of IBSA like sextortion and NCEI. Qin \etal discussed options such as disappearing messages, screenshot notifications, and watermarking to prove ownership or track the origin of a leaked image~\cite{elissa-ibsa-framework}. In practice, the effectiveness of these technologies hinges on the origin of an image (\eg initially consensual vs.\ covertly recorded), and the willingness of perpetrators to adhere to tech-mediated norms. 
However, any increased friction for perpetrators can still reduce harm~\cite{redmiles2024friction}.

\paragraph{Audio-visual alerts around recording} \Vses of NCEI shared that perpetrators often covertly captured explicit imagery. One intervention explored by device manufacturers in Japan and South Korea has been to emit a ``shutter'' sound whenever a cellphone records imagery~\cite{shutter-sound-ncei}. This strategy is similar to the alert emitted by AirTags, intended to prevent stalking~\cite{air-tag-stalking}. Such a feature could be paired with on-device detection of explicit content. Whether such a feature is acceptable to users, and to what degree it discourages NCEI, requires further investigation. 

\paragraph{Preventing generative content} Specific to nonconsensual synthetic explicit imagery, generative AI technologies require safeguards to prevent their use in IBSA. Solutions likely require a combination of preventing harmful model outputs and also detecting synthetic explicit imagery.
\section{Conclusion}
We examined Reddit conversations about IBSA experiences, exploring the (1) types of IBSA for which people sought help, (2) the help they asked for, and (3) the help they received. 
After identifying over 100,000 posts through combined LLM and manual review, we qualitatively analyzed a stratified sample of 261 posts about seven types of IBSA: financial sextortion, nonfinancial sextortion, nonconsensual \textit{synthetic} explicit imagery, pressurized sexting, cyberflashing, nonconsensual explicit imagery, and recorded sexual assault.
We synthesized similarities in types of help sought and given, finding that across the seven IBSA types we studied, \vses most often asked for and were offered information, empathy and therapeutic support, and advice about managing existing relationships.
Technical, legal, and other institutional support were comparatively less common, indicating opportunities for more comprehensive support.
Our work informs existing challenges towards mitigating, preventing, and supporting recovery from IBSA, and we outline the role technology could have towards potential solutions.

\section*{Acknowledgements}
We thank our anonymous reviewers for their valuable feedback and Nicola Henry for reviewing our search keywords.

{\footnotesize
\bibliographystyle{abbrv}
\interlinepenalty=10000
\bibliography{citations}
}

\appendix
\section{Additional Methodological Details}

\footnotesize

\subsection{Subreddits used}
\label{sec:subreddits}

The subreddits we manually selected, as described in Section~\ref{sec:data}, were:
\textsf{r/advice, r/techsupport, 
r/legaladvice, r/legaladviceuk, r/legaladviceindia, r/legaladvicecanada, r/legaladviceeurope, r/legaladvicegerman, r/legaladviceNZ, r/legaladviceireland, r/legaladviceeu, 
r/relationship\_advice, r/dating\_advice, r/relationships, r/dating, r/marriage, r/sex, 
r/online\_dating\_advice, r/indian\_datingadvice, r/femaledatingstrategy, r/onlinedating, 
r/tinder, r/bumble, r/hingeapp, r/grindr, 
r/onlyfansadvice, r/fansly\_advice, r/camgirlproblems, 
r/antipornography, r/loveafterporn, r/pornfreerelationships, r/pornfree, 
r/sexualassault, r/sexualassaultsurvivor, r/creepyPMs, 
r/twoxchromosomes, r/thegirlsurvivalguide, r/askwomen, r/askwomennsfw, r/askmen, r/askgaybros, 
r/sextortion, r/scams}.

\begin{figure}[t]
\centering

\begin{minipage}{\columnwidth}
   \begin{tcolorbox}[colback=blue!5!white,
                    colframe=blue!20!white,
                    arc=4mm,
                    ]
     \scriptsize {\bf Title:} [TITLE]\\
{\bf Article:} [ARTICLE]\\

{\bf Task:} Categorize the Title and Article as either\\
- 1) Sextortion\\
- 2) Nude image taken without consent\\
- 3) Cyberflashing or receiving unwanted nudity\\
- 4) Manipulated media like deepfakes, photoshopped images, or cheapfakes\\
- 5) Pressuring someone into sending nude images\\
- 6) Unwanted attention\\
- 7) Recording of sexual assault\\
- 8) Other\\

Categorization:"""
  \end{tcolorbox}
\end{minipage}
\vspace{-0.5em}
\caption{Categorization prompt designed to identify the most relevant type of IBSA discussed in a Reddit post. IBSA type names and description differ from types shown in Table~\ref{tab:ibsa-categories} due to experimental iteration with LLM prompts.}
\label{fig:ibsa_category_prompt_1}
\end{figure}

\begin{figure}[t]
\centering

\begin{minipage}{\columnwidth}
   
   \begin{tcolorbox}[colback=yellow!5!white,
                    colframe=yellow!20!white,
                    arc=4mm,
                    ]
     \footnotesize 
     
You are a content moderator for safety policies. Your job is to carefully inspect articles and answer each question below using the single word "Yes" or "No".\\

{\bf Title:} [TITLE]\\
{\bf Article:} [ARTICLE]\\

{\bf Question 1:} Does the title or article discuss a threat to expose nude images unless a payment is made?\\
{\bf Question 2:} Does the title or article discuss a threat to expose nude images unless the person sends more nude images or stays in a relationship?\\
{\bf Question 3:} Does the title or article discuss taking, recording, or leaking a nude image without consent?\\

Your answer should be in json format seen below, where your answer should replace the ``` :\\
\{\\
  ``Question 1'': ```,\\
  ``Question 2'': ```,\\
  ``Question 3'': ```\\
\}
  \end{tcolorbox}
\end{minipage}
\vspace{-0.5em}
\caption{Categorization prompt to determine whether a post discussed a specific type of IBSA (here, nonfinancial sextortion). A Python parser validated the sequence of Yes and No answers necessary for a category label to apply.}
\label{fig:ibsa_category_prompt_2}
\end{figure}

\subsection{Filtering: Keywords vs. LLM Prompts}
\label{sec:extra-prompts}
As described in Section~\ref{sec:data}, we initially explored using keywords chosen from prior work, manual searches, discussion with experts, and our own domain expertise.
We collected 60+ keywords and developed a three-part formula for surfacing relevant posts: >1 media keyword (e.g., \textit{image, nude, screenshot, personal vid}), >1 IBSA keyword (e.g., \textit{revenge porn, cyberflashing, without consent, sextortion, leak, threat, coerce}), >1 help keyword (e.g., \textit{help, support, what should i do}).
However, keyword searches introduced a prohibitive number of false positives to manually review, e.g., discussing IBSA in the news, posts that were not help-seeking.

Therefore, we developed the LLM-based filtering and categorizing approach described in Section~\ref{sec:data};
figures~\ref{fig:ibsa_category_prompt_1} and~\ref{fig:ibsa_category_prompt_2} show additional prompts.
All posts were subsequently validated through manual review.

\subsection{Codebook}
\label{sec:codebook}

The full list of codes applied to the 261 posts in our post dataset, and the 160 associated threads, with abbreviated definitions for each code. Codes were not mutually exclusive. \\

\noindent \textbf{1) Codes about the Nature of IBSA}

\noindent \textbf{Perpetrator}: current partner, ex-partner, friend, family members, colleague (work or school), stranger (unknown to \vs prior to IBSA), other (none of the above), unspecified

\noindent \textbf{Perpetrator(s)' and \vses' gender, age, location}: in terms used by post

\noindent \textbf{Method of distribution}
\begin{itemize}
    \setlength\itemsep{0em}
    \item Threat: Perpetrator threatens to share or distribute the image
    \item Possession: Perpetrator has possession of the image without threatening to or actually sharing
    \item Social media: Perpetrator posted the image to social media
    \item Website: Perpetrator posted the image to a website or elsewhere online
    \item Messaging: Perpetrator directly messaged the image to others
    \item N/A (no content): Perpetrator pressured the VS to send an image, but VS did not
    \item Unspecified
\end{itemize}

\noindent \textbf{Image origin}
\begin{itemize}
    \setlength\itemsep{0em}
    \item Consented: Shared for a different reason or in a prior context, but consent has been revoked 
    \item Coerced: Sent or taken under coercion from the perpetrator including:  under false pretenses; non-secret, non-consensual image taking; soliciting face images for abuse
    \item Coerced (not sent): VS has not sent image, but perpetrator attempted coercion
    \item Created: Synthetically created
    \item Hacked: Obtained through hacking or unauthorized access
    \item Sold: Available for sale by VS, but not for sharing
    \item Secretly recorded: Recording not known to VS when recorded 
    \item Unconsented exposure: VS nonconsensually exposed to someone else's image
    \item Unspecified
\end{itemize}

\noindent \textbf{2) Codes about Help-Seeking}

\noindent \textbf{Self-help}: Yes (poster is the person experiencing the abuse),
No (poster is NOT the person experiencing the abuse)

\noindent \textbf{Strategies attempted}
\begin{itemize}
    \setlength\itemsep{0em}
    \item Support: Talking with others in a support network, online or offline; does not include the Reddit post itself
    \item Platform report: Using online platform reporting mechanisms
    \item Police report: Contacting the police or making a police report
    \item Workplace report: Contacting HR or workplace supervisor
    \item Engage: Negotiating conditions of abuse with the perpetrator
    \item Disengage: Stopping contact or other means of engagement with the perpetrator to try to mitigate the abuse
    \item Securing accts: Taking actions to secure online accounts
    \item Evidence: Recording evidence of the abuse
    \item Delete content: Deleting images, even if not all copies
    \item Other
    \item Unspecified
\end{itemize}
\noindent 

\noindent \textbf{Help sought}: See Table~\ref{tab:help-sought}.

\noindent \textbf{When poster sought help}:
Based on the user states framework~\cite{jigsawmedium}.
\begin{itemize}
    \setlength\itemsep{0em}
    \item Prevention: Aiming to prevent future exposure to digital-safety risks; often but not always during relatively calm state of mind.
    \item Monitoring: Watching for digital-safety events to quickly respond; often with low to moderate stress.
    \item Crisis / Active event: Actively experiencing a digital-safety event (possibly hours/days); likely want to stop event; often very high stress.
    \item Recovery: Digital-safety event stopped or it's been determined stopping it may not be possible; ready to address damage; possibly much later; often moderate to high stress.
    \item Multiple events: Complex cases with multiple or overlapping digital-safety events; often heightened stress and trauma.
\end{itemize}

\noindent \textbf{3) Codes about Help-Giving}:
See Table~\ref{tab:help-given}.

\section{Additional Results}
\label{sec:extra-results}

Tables~\ref{tab:strategies-attempted} and~\ref{tab:platforms} show, respectively and for each IBSA type, strategies previously attempted by \vses and platform(s) involved.

\begin{table}[h]
\scriptsize
\centering
\begin{tabular}{lrrrrrrr}
\toprule
    \textbf{Strategy}
    &\rotatebox{90}{\textbf{FS}} & \rotatebox{90}{\textbf{NFS}} & \rotatebox{90}{\textbf{NCSEI}} & \rotatebox{90}{\textbf{PS}} & \rotatebox{90}{\textbf{CF}} & \rotatebox{90}{\textbf{NCEI}} & \rotatebox{90}{\textbf{RSA}} \\ 
\midrule
Disengage & 22 & 18 & 22 & 18 & 10 & 8 & 12 \\
Engage & 16 & 10 & 11 & 22 & 8 & 15 & 6 \\
Securing accounts & 22 & 9 & 9 & 0 & 2 & 2 & 0 \\
Social support & 5 & 5 & 2 & 2 & 0 & 7 & 5 \\
Police report & 10 & 0 & 4 & 0 & 2 & 4 & 5 \\
Platform report & 9 & 1 & 6 & 1 & 1 & 2 & 1 \\
Evidence & 7 & 0 & 6 & 0 & 1 & 2 & 1 \\
Delete content & 2 & 1 & 0 & 1 & 0 & 4 & 2 \\
Workplace report & 0 & 1 & 1 & 2 & 4 & 0 & 0 \\
\midrule
Other & 2 & 1 & 2 & 0 & 1 & 1 & 0 \\
Unspecified & 9 & 14 & 13 & 8 & 17 & 18 & 17 \\
\bottomrule
\end{tabular}
\vspace{-0.5em}
\caption{Strategies attempted before posting on Reddit.}
\label{tab:strategies-attempted}
\end{table}

\begin{table}[h]
\scriptsize
\centering
\begin{tabular}{lrrrrrrr}
\toprule
\textbf{Platform} & \rotatebox{90}{\textbf{FS}} & \rotatebox{90}{\textbf{NFS}} & \rotatebox{90}{\textbf{NCSEI}} & \rotatebox{90}{\textbf{PS}} & \rotatebox{90}{\textbf{CF}} & \rotatebox{90}{\textbf{NCEI}} & \rotatebox{90}{\textbf{RSA}} \\
\midrule
Messaging (apps, DMs) & 42 & 26 & 30 & 31 & 32 & 15 & 8 \\
Mainstream social media & 31 & 16 & 23 & 3 & 5 & 10 & 4 \\
Device & 2 & 11 & 9 & 3 & 0 & 18 & 27 \\
Dating app & 11 & 2 & 5 & 7 & 6 & 0 & 0 \\
Adult content (websites, apps) & 0 & 1 & 2 & 0 & 0 & 3 & 4 \\
Social discovery app & 1 & 1 & 0 & 0 & 1 & 2 & 0 \\
Financial / commerce & 3 & 0 & 1 & 0 & 1 & 1 & 0 \\
Website & 0 & 1 & 0 & 0 & 0 & 2 & 0 \\
\midrule
Unspecified & 0 & 4 & 1 & 7 & 2 & 1 & 1 \\
\bottomrule
\end{tabular}
\vspace{-0.5em}
\caption{Platforms involved in the IBSA. Messaging also includes messages or images otherwise described in posts as ``sent.''} 
\label{tab:platforms}
\end{table}

\end{document}